\documentclass[floatfix,aps,prd,showpacs,amsmath,nofootinbib,preprintnumbers,onecolumn]{revtex4}

\usepackage{graphicx}
\usepackage{bm}

\def\half{{1\over 2}}

\def\p{\partial}

\def\half{{1\over 2}}
\def\({\left(}
\def\){\right)}
\def\[{\left[}
\def\]{\right]}

\def\e{\begin{equation}}
\def\q{\end{equation}}
\def\m{\begin{eqnarray}}
\def\n{\end{eqnarray}}

\begin{document}

\title{Large Local Non-Gaussianity from General Single-field Inflation}

\author{Qing-Guo Huang$^1$ and Yi Wang$^2$}
\email{huangqg@itp.ac.cn, yi.wang@ipmu.jp}
\affiliation{$^1$ Kavli Institute for Theoretical Physics China, State Key Laboratory of Theoretical Physics, Institute of Theoretical Physics, Chinese Academy of Science, Beijing 100190, People's Republic of China \\
$^2$ Kavli Institute for the Physics and Mathematics of the Universe,
Todai Institutes for Advanced Study, University of Tokyo,
5-1-5 Kashiwanoha, Kashiwa, Chiba 277-8583, Japan}

\date{\today}

\begin{abstract}

We investigate the non-Gaussian signatures of ultra slow-roll inflation. The bispectrum and the trispectrum are calculated with general initial conditions. The trispectrum  is of local shape, as in the case of the bispectrum. We show that the prediction of local non-Gaussianity is robust again generalizing the Bunch-Davies vacuum. The Suyama-Yamaguchi relation is saturated in this scenario.

\end{abstract}


\maketitle


Inflation \cite{Guth:1980zm, Linde:1981mu, Albrecht:1982wi} generates remarkably Gaussian fluctuations \cite{Mukhanov:1981xt}. While the departure from the Gaussianity plays a key role in classifying inflation models \cite{Chen:2010xka, Wang:2013zva}. The non-Gaussianities generated in the primordial epoch can be characterized by their shapes. For example, the local shape non-Gaussianity \cite{Komatsu:2001rj, Maldacena:2002vr} indicates large super-Hubble interactions; equilateral \cite{Chen:2006nt} or orthogonal \cite{Creminelli:2005hu} shapes indicate modifications of the kinetic Lagrangian of the inflaton; intermediate shapes \cite{Chen:2009we, Chen:2009zp} are signs of existence for interacting sectors with energy scale of order Hubble; and a folded shape \cite{Chen:2006nt} arises from modified initial conditions. 

To relate inflationary models to observations, on the one hand, it is important to find out which model better fits data. On the other hand, it is equally (if not more) important to be able to rule out classes of models, with general assumptions. Consistency relations for non-Gaussianities are here in position for the ability to rule out models. 

The best studied consistency relation for non-Gaussianity is the Maldacena's squeezed limit \cite{Maldacena:2002vr, Creminelli:2004yq, Li:2008gg}: In a $3$-point function  $\langle\zeta_{\mathbf{k}_1}\zeta_{\mathbf{k}_2}\zeta_{\mathbf{k}_3}\rangle$, when one of the mode has wavelength much greater than the other two (say, $k_1 \ll k_2 \simeq k_3$), at the horizon crossing time of $k_2$ and $k_3$, the mode $k_1$ is already super-Hubble and thus behaves as a shift of background. For single field inflation, this shift of background can be characterized fully by a modified Hubble crossing time for mode $k_2$ and $k_3$. As a result, the $3$-point function  $\langle\zeta_{\mathbf{k}_1}\zeta_{\mathbf{k}_2}\zeta_{\mathbf{k}_3}\rangle$ can be calculated from the scale dependence of the power spectrum as
\begin{align}\label{eq:mald}
  \lim_{k_1/k_3\rightarrow0} \langle\zeta_{\mathbf{k}_1}\zeta_{\mathbf{k}_2}\zeta_{\mathbf{k}_3}\rangle 
  = - (2\pi)^7 \delta^3(\mathbf{k}_1+\mathbf{k}_2+\mathbf{k}_3)
  \frac{P_\zeta^2}{4k_1^3k_3^3} (n_s-1) ~.
\end{align}
This consistency relation can be understood as a no-go theorem, obstructing single field inflation to produce large local non-Gaussianities.

However, no-go theorems are no better than their assumptions. When assuming the long wave length mode behaves as a shift of background, one need to be careful about the differences between a long wave length mode and a shift of background. The differences include:
\begin{itemize}
\item The long wave length mode exits the horizon at an earlier time during inflation. However, a shift of background never exits the horizon. Thus initial correlations between $\mathbf{k}_1$ and $\mathbf{k}_2$, $\mathbf{k}_3$, if not suppressed in the $k_1/k_2 \rightarrow 0$ limit, may provide a violation of the no-go theorem. There is indeed known counter example of this type: by allowing  non-Bunch-Davies initial conditions for inflation \cite{Chen:2006nt}.
\item The dynamics of the background scale factor is governed by the Friedmann equation, which is a first order differential equation, with only one mode (i.e. only one integration constant). However, the long wave length perturbation satisfies a second order differential equation, and thus has two modes. On inflationary background typically one of the two perturbation modes decays exponentially. Thus only one constant mode is the leftover and can be matched to a shift of background. However, counter examples are also be known of this type. For example, in bouncing cosmology \cite{Cai:2009fn}, the comoving curvature perturbation is growing on super-Hubble scales and the consistency relation is violated.

Recently, it is noticed that the growing mode of comoving curvature perturbation can play a role during inflation as well \cite{Namjoo:2012aa, Martin:2012pe}. This is because the comoving curvature perturbation satisfies the following equation of motion (EoM):
\begin{align} 
  \ddot \zeta_k + (3+\eta) H \dot \zeta_k + \frac{c_s^2k^2}{a^2} \zeta_k = 0~.
\end{align}
Note that not only the Hubble friction $3H$ is present in the EoM, but there is also a term $\eta H$ multiplying $\dot\zeta_k$. In case that $\eta<-3$, the Hubble friction turns into a boost, and the conventionally decaying mode becomes growing. The possibility of $\eta<-3$ can be realized by letting the inflaton climb up the potential using its kinetic energy. It is shown that when $\eta\simeq -6$, the power spectrum is still nearly scale invariant. But the Maldacena's relation for non-Gaussianity \eqref{eq:mald} is violated. This scenario is dubbed as ultra slow-roll inflation. In \cite{Namjoo:2012aa, Martin:2012pe}, a standard kinetic term is considered and the local non-Gaussianity is still $f_\mathrm{NL}\sim 1$. On the other hand, one can consider the single field inflation with small sound speed \cite{Chen:2013aj}. In this specific model the local non-Gaussianity is boosted into $f_\mathrm{NL}\sim 1/c_s^2$.
\end{itemize}

In this paper we investigate in detail the ultra slow-roll inflation scenario in terms of general single field inflation. Let's start with the action 
\m
S=\half \int d^4 x\sqrt{-g} \[M_p^2 R+2 P(X,\phi)\], 
\n
where $\phi$ is the inflaton and $X\equiv -\half g^{\mu\nu}\p_\mu\phi\p_\nu\phi$. 
The energy density of inflaton field is 
\m
\rho=2XP_{,X}-P, 
\n
where $P_{,X}$ denotes the derivative with respect to $X$. The dynamics of the universe is govern by 
\m
H^2&=&{\rho\over 3M_p^2}, \\
\dot \rho&=&-3H(\rho+P). \label{drho}
\n

For convenience, we introduce the sound speed $c_s$ as 
\m
c_s^2\equiv {dP\over d\rho}={P_{,X}\over P_{,X}+2XP_{,XX}}, 
\n
and slow-roll parameters 
\m
\epsilon&=&-{\dot H\over H^2}={XP_{,X}\over M_p^2H^2}, \\
\eta&=&{\dot \epsilon \over H\epsilon}. 
\n
From the above two equations, we obtain
\m
\eta=2\epsilon+{P_{,X}+XP_{,XX}\over P_{,X}} {\dot X\over HX}+{P_{,X\phi}\over P_{,X}}{\dot \phi\over H}.
\n
From Eq.~(\ref{drho}), 
\m
{1\over c_s^2} {\dot X\over HX}=-{2XP_{,X\phi}-P_{,\phi}\over XP_{,X}}{\dot \phi\over H}-6~, 
\n
and then 
\m
\eta=2\epsilon-6{P_{,X}+XP_{,XX}\over P_{,X}+2XP_{XX}}+{-XP_XP_{,X\phi}+(P_{,X}+XP_{,XX})P_{,\phi}\over XP_{,X}(P_{,X}+2XP_{,XX})} {\dot \phi\over H}~. 
\n
In the limit of $c_s\ll 1$, i.e. $XP_{,XX}\gg P_{,X}$,
\m
\eta\simeq 2\epsilon-3+\left(
  -\frac{P_{,X\phi}}{2XP_{,XX}} + \frac{P_\phi}{2XP_{,X}}
\right){\dot \phi\over H}~. 
\n

To calculate the perturbations, we decompose the metric as
\begin{align}
  ds^2 = - N^2dt^2 + h_{ij} (dx^i + N^i dt) (dx^j + N^j dt)~,
\end{align}
where for scalar perturbations, we have $h_{ij} =  a^2 e^{2\zeta}\delta_{ij}$. After solving the constraints, the second order gravitational action is
\begin{align}
  S = \int d^3x~dt~ M_p^2 \epsilon \left(  \frac{a^3}{c_s^2} \dot\zeta^2 - a(\partial\zeta)^2 \right)~.
\end{align}
The linear perturbation can be solved and quantized as
\begin{align}
  \zeta_\mathbf{k} = u_k a_\mathbf{k} + u^*_k a^\dagger_{-\mathbf{k}}~,
\end{align}
where $u_k$ satisfies the linearized equation of motion
\begin{align} \label{eq:ueom}
  \ddot u_k + (3+\eta) H \dot u_k + \frac{c_s^2k^2}{a^2} u_k = 0~.
\end{align}
When $\eta<-3$, the $\zeta$ field effectively lives in a contracting universe (although the real scale factor is expanding). Thus the growing mode dominates over the constant mode. Here we shall focus on the case $\eta\simeq -6$, in which the density perturbation has a nearly flat spectrum. The solution of \eqref{eq:ueom} is
\begin{align}
  u_k = \frac{H}{2M_p\sqrt{\epsilon_i c_s k^3}}  
  \left( \frac{\tau_i}{\tau} \right)^3 (-1-ikc_s\tau) e^{-ikc_s\tau}~.
\end{align}
The power spectrum for density fluctuations can be calculated as
\m
P_R={H^2\over 8\pi^2 M_p^2\epsilon_ic_s}\({\tau_i\over \tau_e}\)^6 ~. 
\n
Note that the kinetic energy of inflaton will eventually be used up and then inflation transits into the conventional slow-roll era with $|\eta|\ll 1$. The transition time would either correspond to comoving scales of CMB observations (about the largest 10 e-folds of observable cosmological scales) or smaller scales. In case that the transition happens at the CMB scales, extra features should be expected on the power spectrum, and one cannot trust the extremely squeezed limit for non-Gaussianities. On the other hand, in order that this ultra slow-roll epoch lasts longer than 10 e-folds, one requires $\epsilon$ to be exponentially decaying throughout this period. This results in inflation with very low energy scales, $H< 10$~GeV.

For non-linear perturbations, we can alternatively work in the $\delta\phi$-gauge, where the inflaton $\phi$ has fluctuation $\delta\phi$, on a spatially flat slice $\zeta=0$. After perturbing the action, we do a gauge transformation
\begin{align}
  \zeta_k = - \frac{H}{\dot\phi} \delta\phi_k + \mathcal{O}(\delta\phi^2)
\end{align}
to go back to the $\zeta$ gauge. Note that the $\mathcal{O}(\delta\phi^2)$ terms do not contribute to the leading order expansion in the small $c_s$ limit.

For the growing mode, the time derivative $\dot\zeta_k$ in the Hamiltonian is more important than the spatial derivative $(k/a)\times \zeta_k$. Thus we can further neglect the spatial derivatives. As a result, the leading terms in the third order and fourth order Hamiltonian are respectively \cite{Chen:2006nt, Chen:2009bc}
\begin{align}
    \mathcal{H}_3 = \frac{2a^3\lambda \dot\zeta^3}{H^3} 
  = 2 \frac{M_p^2 \epsilon_i}{H c_s^2} \frac{\lambda}{\Sigma} \left( \frac{\tau}{\tau_i}  \right)^6 (\zeta')^3 ~,
\end{align}
and
\begin{align}
  \mathcal{H}_4 = \frac{a^3}{H^4} \left( -\mu + 9 \frac{\lambda^2}{\Sigma} \right) \dot\zeta^4
  = \frac{M_p^2\epsilon_i}{aH^2c_s^2}  \left( \frac{\tau}{\tau_i}  \right)^6
  \left( -\frac{\mu}{\Sigma} + 9 \frac{\lambda^2}{\Sigma^2}  \right) (\zeta')^4 ~.
\end{align}

To calculate correlation functions, we insert the above interaction Hamiltonian into the in-in formalism. At first and second order, the in-in formalism can be written as
\begin{align}\label{eq:inin1}
  \langle\Omega|Q(\tau)|\Omega\rangle_1 = 2\mathrm{Im} \int_{\tau_0}^{\tau} d\tau_1 \langle0| Q^I(\tau) H_I(\tau_1) |0\rangle~,
\end{align}
\begin{align}\label{eq:inin2}
  \langle\Omega|Q(\tau)|\Omega\rangle_2 = \int_{\tau_0}^\tau d\tau_1 \int_{\tau_0}^\tau d\tau_2 \langle0| H_I(\tau_1) Q^I(\tau) H_I(\tau_2) |0\rangle
  - 2\mathrm{Re} \int_{\tau_0}^\tau d\tau_1 \int_{\tau_0}^{\tau_1}d\tau_2 \langle0| Q^I(\tau) H_I(\tau_1) H_I(\tau_2) |0\rangle ~.
\end{align}

Because of the growth of super-Hubble perturbations, the local shape dominates exponentially over the equilateral shape. And the estimators of local shape non-Gaussianity can be calculated as
\begin{align}
  f_{NL} = \frac{15}{2}  \frac{\lambda}{\Sigma} ~, \quad
  g_{NL} = \frac{25}{12} \left( -\frac{\mu}{\Sigma} + 9 \frac{\lambda^2}{\Sigma^2}  \right)~,
  \quad
  \tau_{NL} = 81 \left( \frac{\lambda}{\Sigma} \right)^2~.
\end{align}
Note that the Suyama-Yamaguchi relation \cite{Suyama:2007bg} is saturated: $\tau_{NL} = (36/25) f_{NL}^2$.

\begin{figure}[htbp]
  \centering
  \includegraphics[width=0.8\textwidth]{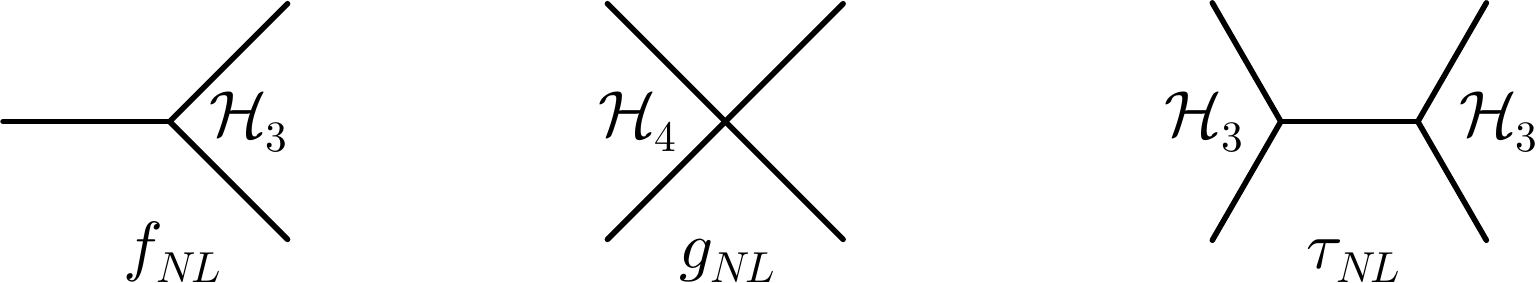}
  \caption{\label{fig:fgt} The contributions to local shape non-Gaussian estimators $f_{NL}$, $g_{NL}$ and $\tau_{NL}$ from the interaction Hamiltonian.}
\end{figure}

The contributions from the interaction Hamiltonian to those estimators are illustrated in Fig.~\ref{fig:fgt}.

In the model of \cite{Chen:2013aj}, $\lambda/\Sigma = 1/(6c_s^2)$, and $\mu/\Sigma = 1/(12c_s^4)$. In this case the non-Gaussian estimators are
\begin{align}
  f_{NL} = \frac{5}{4c_s^2} ~, \quad
  g_{NL} = \frac{25}{72c_s^4} ~, \quad
  \tau_{NL} = \frac{9}{4c_s^4} ~.
\end{align}
One can check the robustness of the above results by considering alternative initial conditions. For this purpose, we consider non-BD vacuum by modifying the mode function into
\m
u_k={H\over M_p \sqrt{4\epsilon_ic_sk^3}}\({\tau_i\over \tau}\)^3 \[C_+(-1-ikc_s\tau)e^{-ikc_s\tau}+C_-(-1+ikc_s\tau)e^{ikc_s\tau}\]. 
\n
The consistency of creation annihilation operators commutation relation, with that between the field and conjugate momentum, gives
\begin{align}
  |C_+|^2-|C_-|^2 = 1~.
\end{align}
The amplitude of primordial power spectrum is 
\m
P_R={H^2\over 8\pi^2M_p^2\epsilon_ic_s}\({\tau_i\over \tau_e}\)^6 |C_++C_-|^2. 
\n
At the level of non-Gaussianity, one can show that
\m
\langle \zeta_{k_1}\zeta_{k_2}\zeta_{k_3}\rangle=(2\pi)^3\delta^{(3)}({\bf k}_1+{\bf k}_2+{\bf k}_3) {9\lambda\over \Sigma}(|C_+|^2-|C_-|^2) (2\pi^2P_R)^2 {\sum_{i=1}^3 k_i^3\over \prod_{i=1}^3 k_i^3}~.
\n
In words, the correction to the three point function is completely written in terms of the correction of $P_R^2$. All the additional corrections can be regrouped into the factor $|C_+|^2-|C_-|^2 = 1$. Therefore, we still have
\m
f_{\rm NL}={15\over 2}  {\lambda\over \Sigma}~. 
\n
Similar conclusion holds for the trispectrum, that although the 4-point correlation is modified, the modification is completely written in terms of $P_R^3$, and the estimator $g_{NL}$ is not modified.

One should note that, as usual, the non-Bunch-Davies vacuum still generates a folded shape of non-Gaussianity. However, the folded shape takes place at the horizon crossing time. Thus the contribution of the folded shape is exponentially small compared with the local shape. Such a folded component in the shape function is not detectable, given enough e-folds of growth for $\zeta_\mathbf{k}$.

To conclude, we calculated the non-Gaussian estimators up to quartic order. The shape of the non-Gaussianity is local and the Suyama-Yamaguchi relation is saturated. Non-Bunch-Davies initial conditions give correction to the correlation functions, but without any change of shape or in terms of the non-Gaussian estimators. 

There are some interesting questions yet to be addressed. For example, it is challenging to derive a generalized consistency relation for single field inflation, which takes this growing mode into account. Also in case the ultra slow-roll inflation period ends when the CMB scales exits the horizon, it is interesting to see what kind of features are imprinted on the power spectrum, as well as to see the squeezed limit with $k_1$ exits the horizon during the ultra slow-roll stage and $k_2$, $k_3$ exits the horizon during the conventional slow roll stage. As another direction, a similar calculation may be done for the generalized Galileons. 

\noindent {\bf Acknowledgments}

QGH is supported by the project of Knowledge Innovation Program of Chinese Academy of Science and a grant from NSFC (grant NO. 10821504). YW is supported by fundings from the Kavli Institute for the Physics and Mathematics of the Universe. YW thanks the University of Cambridge, Institute of Theoretical Physics, the Chinese Academy of Sciences,  and  McGill University for hospitality.




\begin{thebibliography}{99}
\frenchspacing
\bibitem{Guth:1980zm} 
  A.~H.~Guth,
  ``The Inflationary Universe: A Possible Solution to the Horizon and Flatness Problems,''
  Phys.\ Rev.\ D {\bf 23}, 347 (1981).

\bibitem{Linde:1981mu} 
  A.~D.~Linde,
  ``A New Inflationary Universe Scenario: A Possible Solution of the Horizon, Flatness, Homogeneity, Isotropy and Primordial Monopole Problems,''
  Phys.\ Lett.\ B {\bf 108}, 389 (1982).

\bibitem{Albrecht:1982wi} 
  A.~Albrecht and P.~J.~Steinhardt,
  ``Cosmology for Grand Unified Theories with Radiatively Induced Symmetry Breaking,''
  Phys.\ Rev.\ Lett.\  {\bf 48}, 1220 (1982).

\bibitem{Mukhanov:1981xt} 
  V.~F.~Mukhanov and G.~V.~Chibisov,
  ``Quantum Fluctuation and Nonsingular Universe. (In Russian),''
  JETP Lett.\  {\bf 33}, 532 (1981)
  [Pisma Zh.\ Eksp.\ Teor.\ Fiz.\  {\bf 33}, 549 (1981)].

\bibitem{Chen:2010xka} 
  X.~Chen,
  ``Primordial Non-Gaussianities from Inflation Models,''
  Adv.\ Astron.\  {\bf 2010}, 638979 (2010)
  [arXiv:1002.1416 [astro-ph.CO]].

\bibitem{Wang:2013zva} 
  Y.~Wang,
  ``Inflation, Cosmic Perturbations and Non-Gaussianities,''
  arXiv:1303.1523 [hep-th].

\bibitem{Komatsu:2001rj} 
  E.~Komatsu and D.~N.~Spergel,
  ``Acoustic signatures in the primary microwave background bispectrum,''
  Phys.\ Rev.\ D {\bf 63}, 063002 (2001)
  [astro-ph/0005036].

\bibitem{Maldacena:2002vr} 
  J.~M.~Maldacena,
  ``Non-Gaussian features of primordial fluctuations in single field inflationary models,''
  JHEP {\bf 0305}, 013 (2003)
  [astro-ph/0210603].

\bibitem{Chen:2006nt} 
  X.~Chen, M.~-x.~Huang, S.~Kachru and G.~Shiu,
  ``Observational signatures and non-Gaussianities of general single field inflation,''
  JCAP {\bf 0701}, 002 (2007)
  [hep-th/0605045].

\bibitem{Creminelli:2005hu} 
  P.~Creminelli, A.~Nicolis, L.~Senatore, M.~Tegmark and M.~Zaldarriaga,
  ``Limits on non-gaussianities from wmap data,''
  JCAP {\bf 0605}, 004 (2006)
  [astro-ph/0509029].

\bibitem{Chen:2009we} 
  X.~Chen and Y.~Wang,
  ``Large non-Gaussianities with Intermediate Shapes from Quasi-Single Field Inflation,''
  Phys.\ Rev.\ D {\bf 81}, 063511 (2010)
  [arXiv:0909.0496 [astro-ph.CO]].

\bibitem{Chen:2009zp} 
  X.~Chen and Y.~Wang,
  ``Quasi-Single Field Inflation and Non-Gaussianities,''
  JCAP {\bf 1004}, 027 (2010)
  [arXiv:0911.3380 [hep-th]].

\bibitem{Creminelli:2004yq} 
  P.~Creminelli and M.~Zaldarriaga,
  ``Single field consistency relation for the 3-point function,''
  JCAP {\bf 0410}, 006 (2004)
  [astro-ph/0407059].

\bibitem{Li:2008gg} 
  M.~Li and Y.~Wang,
  ``Consistency Relations for Non-Gaussianity,''
  JCAP {\bf 0809}, 018 (2008)
  [arXiv:0807.3058 [hep-th]].

\bibitem{Cai:2009fn} 
  Y.~-F.~Cai, W.~Xue, R.~Brandenberger and X.~Zhang,
  ``Non-Gaussianity in a Matter Bounce,''
  JCAP {\bf 0905}, 011 (2009)
  [arXiv:0903.0631 [astro-ph.CO]].

\bibitem{Namjoo:2012aa} 
  M.~H.~Namjoo, H.~Firouzjahi and M.~Sasaki,
  ``Violation of non-Gaussianity consistency relation in a single field inflationary model,''
  arXiv:1210.3692 [astro-ph.CO].
  
\bibitem{Martin:2012pe} 
  J.~Martin, H.~Motohashi and T.~Suyama,
  ``Ultra Slow-Roll Inflation and the non-Gaussianity Consistency Relation,''
  Phys.\ Rev.\ D {\bf 87}, 023514 (2013)
  [arXiv:1211.0083 [astro-ph.CO]].

\bibitem{Chen:2013aj} 
  X.~Chen, H.~Firouzjahi, M.~H.~Namjoo and M.~Sasaki,
  ``A Single Field Inflation Model with Large Local Non-Gaussianity,''
  arXiv:1301.5699 [hep-th].
  
\bibitem{Chen:2009bc} 
  X.~Chen, B.~Hu, M.~-x.~Huang, G.~Shiu and Y.~Wang,
  ``Large Primordial Trispectra in General Single Field Inflation,''
  JCAP {\bf 0908}, 008 (2009)
  [arXiv:0905.3494 [astro-ph.CO]].

\bibitem{Suyama:2007bg} 
  T.~Suyama and M.~Yamaguchi,
  ``Non-Gaussianity in the modulated reheating scenario,''
  Phys.\ Rev.\ D {\bf 77}, 023505 (2008)
  [arXiv:0709.2545 [astro-ph]].


\end{thebibliography}
\end{document}